\documentclass[useAMS,usenatbib]{mn3e}
\usepackage[T1]{fontenc}
\usepackage{ae,aecompl}

\RequirePackage[colorlinks,citecolor=blue,urlcolor=blue]{hyperref}

\usepackage{amsmath, amssymb}
\usepackage{algpseudocode}
\usepackage{algorithm}
\usepackage[dvips]{graphicx}
\usepackage{subfigure}
\usepackage{lineno}
\usepackage{afterpage}
\usepackage{comment}
\usepackage{enumitem}
\usepackage{multirow}

\let\hat\widehat

\newcommand\hMsun{M_\odot/h}

\usepackage{pbox}

\title[Galaxy-Filament Alignment]{
Investigating Galaxy-Filament Alignments in Hydrodynamic Simulations using Density Ridges}
\author[Yen-Chi Chen et al.]{Yen-Chi Chen,$^{1,3}$\thanks{E-mail:
yenchic@andrew.cmu.edu}
Shirley Ho,$^{2,3}$
Ananth Tenneti,$^{2,3}$
Rachel Mandelbaum,$^{2,3}$\newauthor
Rupert Croft,$^{2,3}$ 
Tiziana DiMatteo,$^{2,3}$
Peter E. Freeman,$^{1,3}$\newauthor
Christopher R. Genovese,$^{1,3}$
Larry Wasserman$^{1,3}$
\\
$^{1}$Department of Statistics, Carnegie Mellon University, Pittsburgh, PA 15213, USA\\
$^{2}$Department of Physics, Carnegie Mellon University, Pittsburgh, PA 15213, USA\\
$^{3}$McWilliams Center for Cosmology, Department of Physics, Carnegie Mellon University, Pittsburgh, PA 15213, USA}
\begin{document}


\pagerange{\pageref{firstpage}--\pageref{lastpage}} \pubyear{2015}

\maketitle

\label{firstpage}

\begin{abstract}

In this paper, we study the filamentary structures and the galaxy alignment 
along filaments at redshift $z=0.06$ in the MassiveBlack-II simulation, a state-of-the-art,
high-resolution hydrodynamical cosmological simulation which includes stellar 
and AGN feedback in a volume of (100 Mpc$/h$)$^3$.
The filaments are constructed using the subspace constrained mean shift 
(SCMS; \citet{Ozertem2011} and
\citet{2015arXiv150105303C}). 
First, we 
show that 
reconstructed filaments using galaxies and reconstructed filaments using dark matter particles
are similar to each other; 
over $50\%$ of the points on the galaxy filaments
have a corresponding point on the dark matter filaments 
within distance $0.13$ Mpc$/h$ (and vice versa)
and this distance is even smaller at high-density regions. 
Second, we observe the alignment of the major principal axis of a galaxy with respect 
to the orientation of its nearest filament
and detect a $2.5$ Mpc$/h$ critical radius for filament's influence on the alignment
when the subhalo mass of this galaxy is between $10^9M_\odot/h$ and $10^{12}M_\odot/h$.
Moreover, we find the alignment signal to increase significantly with the subhalo mass.
Third, 
when a galaxy is close to filaments (less than $0.25$ Mpc$/h$), 
the galaxy alignment toward the nearest galaxy group
depends on the galaxy subhalo mass. 
Finally, 
we find that galaxies close to filaments or groups tend to be rounder
than those away from filaments or groups.
\end{abstract}


\section{Introduction}

Weak gravitational lensing by large-scale structure of the Universe, commonly known as cosmic shear, is a promising technique to constrain cosmological parameters. It is among the key science cases of various ongoing and upcoming surveys, such as the Kilo-Degree Survey (KiDS\footnote{\url{http://kids.strw.leidenuniv.nl/}}), Hyper Suprime Cam (HSC\footnote{\url{http://www.naoj.org/Projects/HSC/}}),  Dark Energy Survey (DES\footnote{\url{http://www.darkenergysurvey.org/}}), Large Synoptic Survey Telescope (LSST\footnote{\url{http://www.lsst.org/lsst/}}), Euclid\footnote{\url{http://sci.esa.int/euclid/}} and the Wide-Field Infrared Survey Telescope (WFIRST\footnote{\url{http://wfirst.gsfc.nasa.gov/}}). Theoretically, it provides valuable information on both the geometry and structure growth of the Universe without assumptions on the relationship between the luminous and dark matter; see e.g
\cite{2002NewAR..46..767H, 2005A&A...429...75V, 2006ApJ...644...71J, 2010A&A...516A..63S, 
2012ApJ...761...15L, 2012MNRAS.427..146H, 2013ApJ...765...74J, 2014MNRAS.440.1322H}. 
If systematics can be sufficiently controlled for future missions, cosmic shear has the potential to be the most constraining cosmological probe, thus understanding and controlling systematics in the weak lensing measurements needs to improve significantly over the next few years for its potential to come to fruition with large data volumes in these upcoming surveys.

Cosmic shear is typically measured through two-point correlations of observed galaxy ellipticities. In the weak lensing regime, the observed ellipticity of a galaxy is the sum of its intrinsic ellipticity $\epsilon_t$ and gravitational shear $\gamma: \epsilon_{\sf obs} \approx \epsilon_t   + \gamma$. When the intrinsic shapes of galaxies are spatially correlated, these intrinsic alignment (IA) correlations can contaminate the gravitational shear signal and lead to biased measurements if not properly removed or modeled. Since early work establishing its potential effects
\citep{2000ApJ...545..561C, 2000MNRAS.319..649H, 2001MNRAS.320L...7C, 2001ApJ...559..552C}, 
IA has been examined through observations 
(e.g., \citealt{2007MNRAS.381.1197H, 2011A&A...527A..26J, 2012JCAP...05..041B, 
2014AAS...22442304S}), analytic modeling, and simulations 
(e.g., \citealt{2012JCAP...05..030S,2014MNRAS.441..470T,2015MNRAS.448.3522T}).
See \cite{2014PhRvD..89f3528T, 2015arXiv150405456J}, 
and references therein, for recent reviews.

To create a fully predictive model of IA would require the understanding of the complex processes involved in the formation and evolution of galaxies and their dark matter halos, as well as how these processes couple to large-scale environment. The shapes of elliptical, pressure-supported galaxies are often assumed to align with the surrounding dark matter halos, which are themselves aligned with the stretching axis of the large-scale tidal field 
\citep{2001MNRAS.320L...7C, 2004PhRvD..70f3526H, 2010PhRvD..82d9901H}. 
This tidal alignment model leads to shape alignments that scale linearly with fluctuations in the tidal field, and it is thus sometimes referred to as ``inear alignment'', although nonlinear contributions may still be included
\citep{2007NJPh....9..444B, 2011JCAP...05..010B, 2015arXiv150402510B}. As for spiral galaxies, where angular momentum is likely the primary factor in determining galaxy orientation, IA modeling is typically based on tidal torquing theory, leading to a quadratic dependence on tidal field fluctuations
\citep{2000ApJ...543L.107P, 2001MNRAS.320L...7C, 2002astro.ph..5512H, 2008ApJ...686L...1L}, 
although on sufficiently large scales, the contribution linear in the tidal field may become dominant.

Filaments in the cosmic web play a key role in the large- scale tidal field since they are related to the gradient and the Hessian matrix of the matter density field
\citep{2007MNRAS.375..489H,2007MNRAS.381...41H,
2008MNRAS.383.1655S,2009MNRAS.396.1815F, 2010MNRAS.408.2163A, Cautun2012}.
As a result, it is expected that filaments would affect the shapes and alignments of their nearby galaxies. There is evidence for anti-alignment between the galaxy spin orienta- tion of disk galaxies and the direction of nearby filaments
\citep{2013MNRAS.428.2489L, 
2013MNRAS.428.1827T,2013ApJ...775L..42T, 
2014MNRAS.440L..46A, 2014MNRAS.444.1453D,
2014MNRAS.445L..46W, 2015MNRAS.446.2744L, 2015ApJ...798...17Z},
and an alignment between galaxy shape and the orientation of the nearest filament
\citep{2006MNRAS.370.1422A,2009ApJ...706..747Z,2013ApJ...775L..42T, 2013ApJ...779..160Z}.
Other evidence also indicates that 
direction of a galaxy pair \citep{2015A&A...576L...5T}
and the alignment of satellite galaxies toward their host galaxies \citep{2015MNRAS.450.2727T}
are dependent on their nearby filaments.
The effect of filaments on intrinsic galaxy shapes is worth studying since filaments are expected to trace the tidal fields directly and yet the current theoretical model for IA only includes halos but does not explicitly include filaments; see e.g.
 \cite{2010MNRAS.402.2127S}.

One can, in principle, test and constrain IA models and marginalize over free parameters in specific models; this process could be improved in the quasilinear and nonlinear regime via the tighter priors on model parameters that we could obtain with a 3D filament map of the universe which overlaps with sources in the weak lensing fields. 
In this paper, we apply the subspace constrained mean shift (SCMS) filament finder to the source density in a cosmological hydrodynamic simulation to test whether we can use filaments as a tool to understand the relationship of galaxy alignments with the large-scale density field.

We start with a brief introduction about density ridges,
the model for filaments we are using, in {\S} \ref{sec::ridges}
and describe the MassiveBlack-II Simulation we applied
in {\S} \ref{sec::MBII}.
Then we describe our main results in {\S} \ref{sec::results}
and our conclusions in {\S} \ref{sec::conc}.



\section{Density Ridges as Filaments} \label{sec::ridges}

We detect filaments through SCMS
introduced in \cite{Ozertem2011, 2015arXiv150105303C}.
SCMS is a
gradient-based method that detects filaments as
``galaxy density ridges''.

\begin{figure}
\begin{center}
\includegraphics[width=3 in]{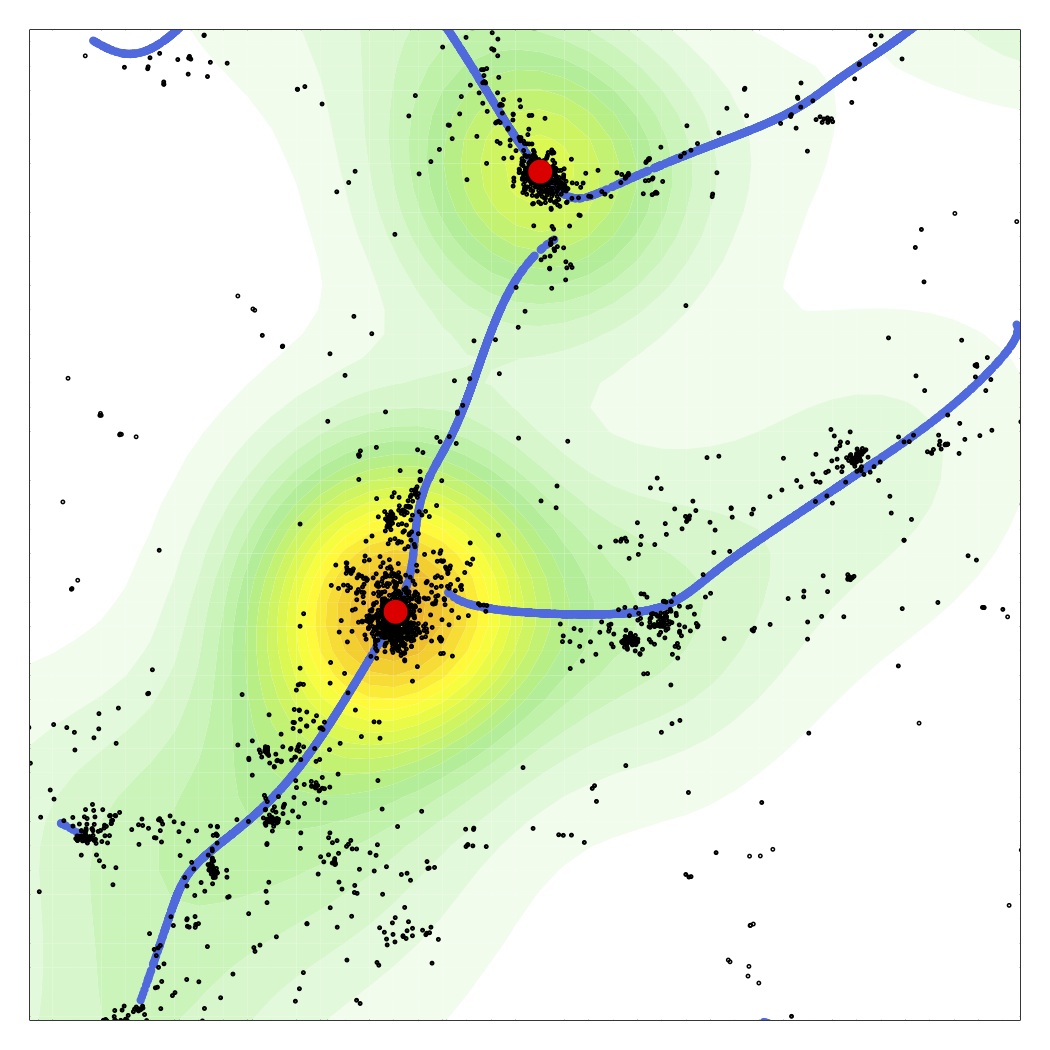}
\caption{Illustration of density ridges (blue curves) in a density field.
Each black dot is a subhalo and the big red dots are galaxy groups.
The color from white-green-yellow-orange denotes the density
induced by smoothing the subhalos.
Note that this 2-D figure is for illustration; our analysis is applied
to 3-D data.
}
\label{Fig::ex::ridge}
\end{center}
\end{figure}

Essentially, filaments from SCMS are defined as follows.
Given galaxies located at $X_1,\cdots,X_n\in\mathbb{R}^3$,
the smoothed density field (also known as the kernel density estimator) is
\begin{equation}
p(x) = \frac{1}{nb^3}\sum_{i=1}^nK\left(\frac{|x-X_i|}{b}\right),
\end{equation}
where $K$ is a Gaussian and $b$ is the smoothing bandwidth
that controls the amount of smoothing on each particle.
In this paper, we apply $b= 1$ Mpc$/h$.
Note that the density field $p$ does not incorporate any 
boundary condition at the edge of the simulation box.
Let $g(x) = \nabla p(x), H(x) = \nabla \nabla p(x)$ be the gradient and 
Hessian matrix of $p(x)$.
Ridges 
of $p(x)$ are the collection of points
\begin{equation}
F = \{x: v_j(x)\cdot g(x)=0, \lambda_2(x)<0, j=2,3\},
\end{equation}
where $v_j(x), \lambda_j(x)$ are the $j$-th eigenvector/value for $H(x)$.
See Figure~\ref{Fig::ex::ridge} for an example about density ridges.
More properties of ridges of a function
can be found in 
\cite{Eberly1996,Ozertem2011,genovese2014nonparametric,2014arXiv1406.1803C,2014arXiv1406.5663C}.

Intuitively, at a point on the ridge,
the gradient is the same as $v_1(x)$, the eigenvector
corresponds to the largest eigenvalue,
and the density curves downward sharply in directions orthogonal to 
the gradient.
When $p$ is smooth and 
the \emph{eigengap}
\begin{align}
\beta(x) = \lambda_1(x) - \lambda_2(x)
\label{eq::eigengap}
\end{align}
is positive,
the ridges have the properties of filaments
\citep{2014arXiv1406.5663C}.
That is, $F$ decomposes into a set of smooth curve-like
structures with high-density.
Since the ridges are smooth, for any point $x\in F$,
the orientation of ridges at $x$, denoted as $\mu_{\sf Ridge}(x)$, is well-defined.

\section{MassiveBlack-II Simulation}	\label{sec::MBII}
In this work, we detect filaments in the MassiveBlack-II (MBII) hydrodynamic simulation \citep{2015MNRAS.450.1349K}, and study the alignments of the shapes of stellar matter components in galaxies with the filaments. MB-II is a state-of-the-art, high-resolution cosmological hydrodynamic simulation performed in a cubic periodic box of size $100$ Mpc$/h$ on a side. The simulations have been performed with the code {\sc p-gadget}, which is a hybrid version of the parallel code {\sc gadget2} \citep{2005MNRAS.361..776S} that has been modified and upgraded to run on Petaflop-scale supercomputers. MBII includes the physics of multiphase interstellar medium (ISM) model with star formation \citep{2003MNRAS.339..289S}, black hole accretion and associated feedback processes \citep{2008ApJ...676...33D,2012ApJ...745L..29D} in addition to gravity and hydrodynamics. This simulation has been run from $z=159$ to $z=0.06$ using $N_\mathrm{part} = 2\times 1792^{3}$ dark matter and gas particles with a gravitational smoothing length, $\epsilon = 1.85$kpc$/h$ in comoving units. The mass of each dark matter particle is $m_\text{DM} = 1.1\times 10^{7}\hMsun$ and the initial mass of a gas particle is
$m_\text{gas} = 2.2\times 10^{6}\hMsun$. The cosmological
parameters in the simulation chosen according to  WMAP7
\citep{2011ApJS..192...18K} are as follows: amplitude of matter
fluctuations $\sigma_{8} = 0.816$, spectral index $n_{s} = 0.96$,
mass density parameter $\Omega_{m} = 0.275$, cosmological constant
density parameter $\Omega_{\Lambda} = 0.725$, baryon density parameter
$\Omega_{b} = 0.046$, and Hubble parameter $h = 0.702$.  

To generate halo catalogs, the friend-of-friends (FOF) halo finder algorithm \citep{1985ApJ...292..371D} is used with a linking length of $0.2$ times the mean interparticle separation. The {\sc
subfind} code \citep{2001MNRAS.328..726S} is used on the halo catalogs to generate subhalo catalogs. Here, subhalos are defined as locally overdense, self-bound particle groups which consist of at least 20 gravitationally bound particles. However, to ensure that the measured shape of a galaxy is reliable, we restrict our galaxy sample to those subhalos with at least $500$ star particles. This is based on the convergence tests in \cite{2014MNRAS.441..470T}. 

\subsection{Shapes of galaxies}\label{shapedef}

In this paper, we use the shapes of the stellar components of galaxies in MBII which are calculated in \cite{2015MNRAS.448.3522T} to quantify intrinsic alignments of galaxies. Here, we give a brief description of the method adopted to calculate these shapes. The shapes of galaxies are modeled as ellipsoids in three dimensions by the eigenvalues and eigenvectors obtained from the eigen-decomposition of the reduced inertia tensor.
We consider all the star particles belonging to the subhalo in the calculation of the inertia tensor. The eigenvectors of the
inertia tensor are denoted as  $\mu_1,\mu_2$, and $\mu_3$ with the
corresponding eigenvalues being
$\lambda_{a},\lambda_{b}$, and $\lambda_{c}$, where $\lambda_{a} >
\lambda_{b} > \lambda_{c}$. The lengths of the principal axes of the ellipsoid
$(a,b,c)$, which are represented by these eigenvectors, are given by square roots of the eigenvalues
$(\sqrt{\lambda_{a}},\sqrt{\lambda_{b}},\sqrt{\lambda_{c}})$. We can now define the intermediate-to-major axis ratio ($q$) and the minor-to-major axis ratio ($s$) as
\begin{equation} \label{eq:axisratios}
q = \frac{b}{a}, \,\, s = \frac{c}{a}
\end{equation}

Note that an iterative procedure is adopted in calculating the shapes such that the axis ratios converge. For additional details regarding the iterative method, see \cite{2015MNRAS.448.3522T}. 
 
For each galaxy, we have two intrinsic quantities: 
the number of star particles $n$ and 
the total mass of this subhalo $M$ (subhalo mass).
Throughout this paper, we use only subhalo mass.

\subsection{Large-Scale Structure}

\begin{table*}
\center
\begin{tabular}{l l c l l}
\hline
Notation & Definition & Type & Remark\\
\hline
$n$& Number of star particles & Integer& $>500$\\
$M$& Halo mass & Scalar&\\
$d_C$& Distance to the nearest galaxy group & Scalar&\\
$d_F$& Distance to the nearest filament & Scalar&\\
$\mu_C$& Direction to the nearest galaxy group & Vector&\\
$\mu_F$& Orientation of the nearest filament & Vector&\\
$q$& First and second principal axis ratio& Scalar& $\in[0,1]$\\
$s$& First and third principal axis ratio & Scalar& $\in[0,1]$\\
$\mu_1$& First principal axis & Vector&\\
$\mu_2$& Second principal axis & Vector&\\
$\mu_3$& Third principal axis & Vector&\\
\hline
\end{tabular}
\caption{Table of quantities associated with a galaxy.}
\label{tab::info}
\end{table*}

\begin{figure}
\begin{center}
\includegraphics{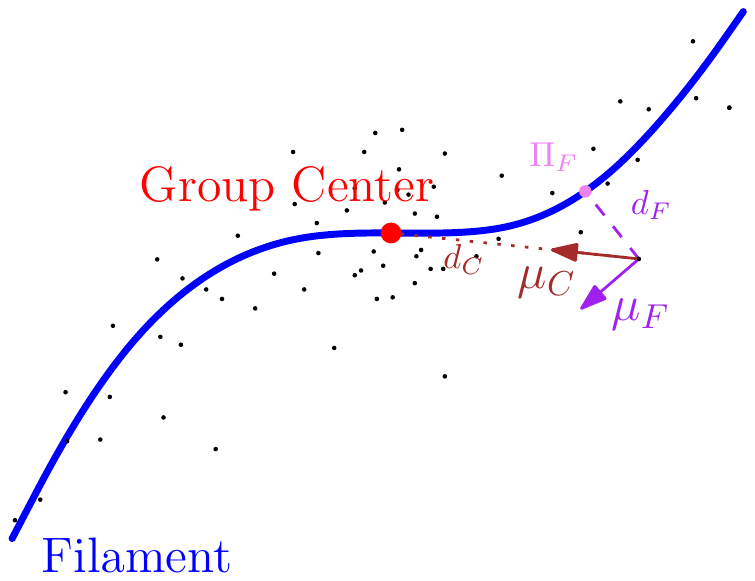}
\caption{Example of $\mu_C,\mu_F, d_C,d_F$. The big red dot denotes
a galaxy group, the blue curve is the filament, and the small black points
are galaxies.
For a given point (galaxy), the brown arrow is $\mu_C$
and the purple arrow is $\mu_F$.
The length of the dotted brown line is $d_C$ and of the dashed purple line is $d_F$.
Note that the projected point onto the filament is the pink point.
}
\label{Fig::ex::CF}
\end{center}
\end{figure}

In our analysis, 
we define galaxy groups as those subhalos with subhalo mass $>10^{13} M_\odot/h$.
This gives $315$ galaxy groups, which corresponds to a comoving number
density $3.15\times 10^{-4}(h/\mbox{Mpc})^3$.
Let $C$ and $F$ denote the collection of all groups (of galaxies) and filaments.
Given any galaxy located at $x$ ($x$ is the $3$D position in the box), 
let $\Pi_C(x)\in C$ and $\Pi_F(x)\in F$
be the `projected' point onto the nearest group/filament.
Namely, $\Pi_C(x)$ (or $\Pi_F(x)$ respectively) is the point in $C$ (or $F$) whose distance to $x$
is smallest among all points in $C$ (or $F$).
In general, this projection is unique 
(although some regions may have non-unique projection, the probability that 
a galaxy falls within these regions is $0$). 
We then define the associated direction to the nearest group and to the nearest filament
as 
\begin{align}
\mu_C(x) &= \frac{\Pi_C(x)-x}{|\Pi_C(x)-x|}\\
\mu_F(x) &= \mu_{\sf Ridge}(\Pi_F(x)).
\end{align}
Note that $\mu_C$ is the direction \emph{toward} the nearest galaxy group
while $\mu_F$ is the direction \emph{of} the nearest filaments.
We also define the projection distances $d_C(x) = |\Pi_C(x)-x|$ and $d_F(x) = |\Pi_F(x)-x|$.
Namely, for a galaxy at position $x$,
$\mu_C(x)$ is the direction from this galaxy to the nearest galaxy group.
And $\mu_F(x)$ is the orientation of the nearby filament.
Figure~\ref{Fig::ex::CF} illustrates all the above quantities in a 2-D picture.

\section{Results}	\label{sec::results}

\subsection{Agreement of Filaments from Galaxies versus Dark Matter Particles}

\begin{figure}
\begin{center}
\includegraphics[width=3 in]{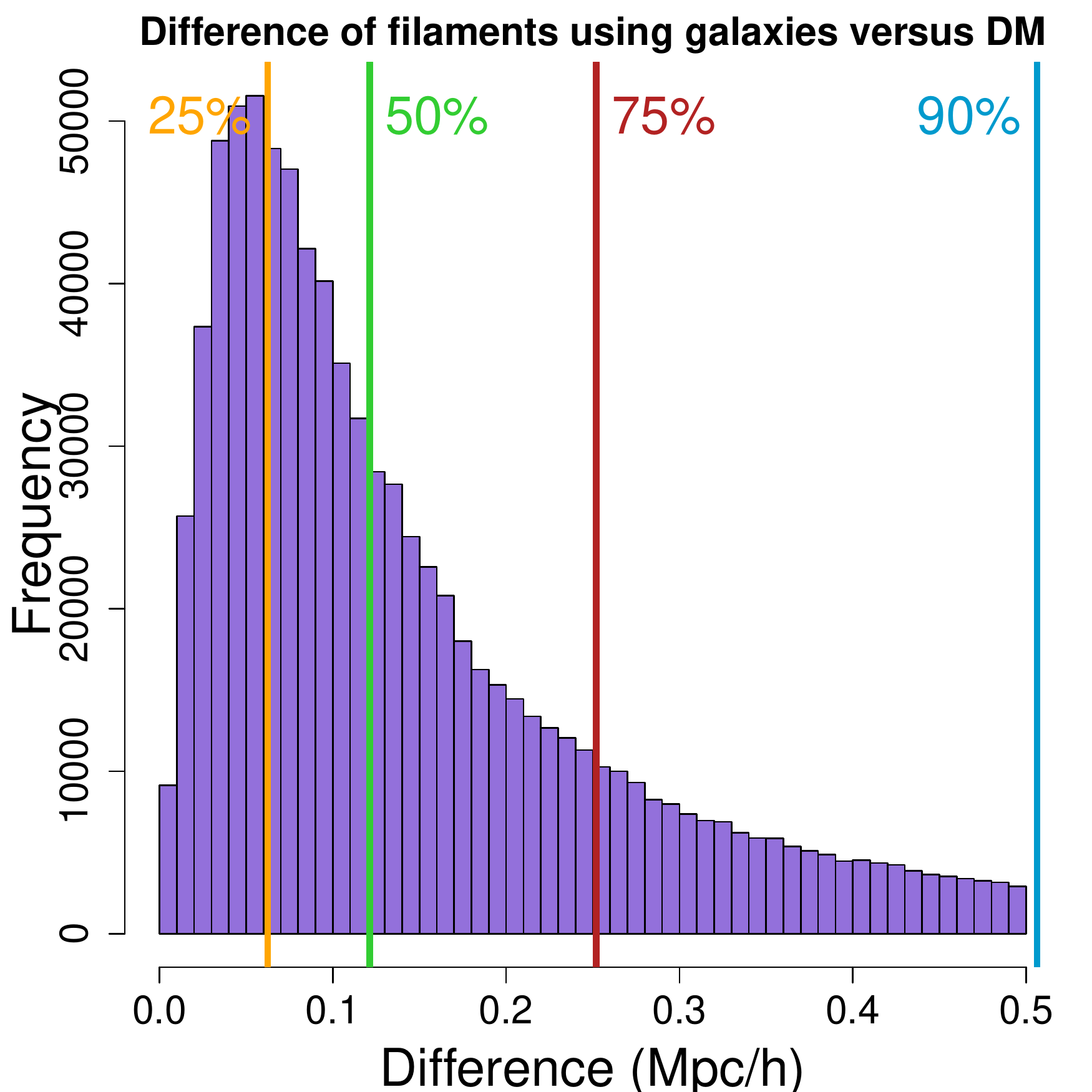}
\caption{Agreement of filaments detected using galaxies versus using dark matter particles.
We apply our filament finder to both galaxies and dark matter particles
and then compare the difference between the two corresponding filament catalogues.
This picture shows the distribution of difference in filament position.
The four color bars denote the quantile values of $25\%, 50\%, 75\%$ and $90\%$.
The difference for filaments recovered from galaxies and dark matter particles is small.
}
\label{Fig::ex::agree}
\end{center}
\end{figure}

To establish the promise of recovering filaments tracing dark matter particles
using only galaxies, 
we first study the similarity for filaments obtained by using galaxies only
versus using dark matter particles. 
The dark matter particles are 
from a random subsample with size $400,000$, which
is approximately the same order as the number of subhalos we have ($\sim 500,000$).
In Figure~\ref{Fig::ex::agree}, we show the difference between the two filament catalogues.
Since both catalogues contain points on the filaments,
the difference is computed using
projected distance between points from two catalogues \citep{2015arXiv150602278C}.
To be more specific, let $\mathcal{F}_D$ and $\mathcal{F}_G$
be the collection of points on the filaments from dark matter particles
and from galaxies.
Then for each point in $\mathcal{F}_D$,
we find its nearest distance to any member of $\mathcal{F}_G$.
This gives a distance value (also known as projected distance) to each point in $\mathcal{F}_D$.
Similarly, for every point in $\mathcal{F}_G$,
we can find the corresponding projected distance to $\mathcal{F}_D$
so that we will assign a distance value to each element of $\mathcal{F}_G$.
The distribution of all the distance values in $\mathcal{F}_D$ and $\mathcal{F}_G$
provides the information about similarity between these two filament catalogues.
If they are similar, this distribution will be concentrated around $0$.
On the other hand, if the two catalogues are different,
this distribution will spread out.
Note that we have tried a random sample with size $800,000$ and the result remains similar.

According to Figure~\ref{Fig::ex::agree},
we see that the two catalogues are similar to each other;
more than 50\% of the points on filaments differ at a distance less than 0.13 Mpc$/h$
and about 90\% of them differ at the distance less than 0.5 Mpc$/h$.

\begin{figure*}
\begin{center}
\includegraphics[width=3 in]{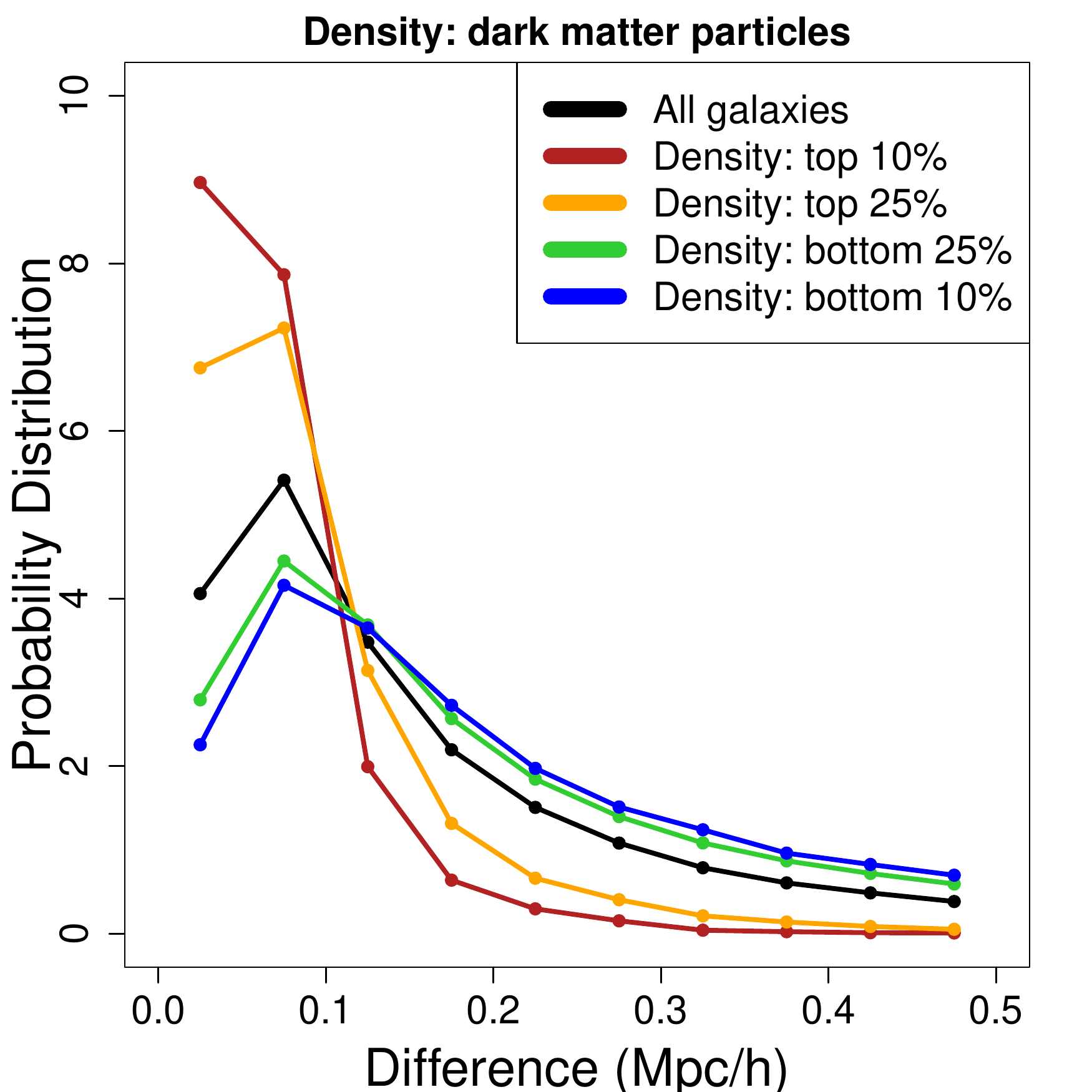}
\caption{Agreement of filaments detected using galaxies versus using dark matter particles
under different environmental densities.
We rank points on filaments according to the densities of dark matter particles
(using densities from galaxies yields a similar result). 
Then we plot distributions of filament differences under different density regimes.
Filaments at high-density regions are colored by red and orange 
(density within the top $10\%$ and $25\%$)
whereas
low density filaments are colored by green and blue
(density in the bottom $25\%$ and $10\%$).
The black curve denotes the distribution using all galaxies 
(which is the distribution of Figure~\ref{Fig::ex::agree}, note that we use the different
bin size).
We see a clear pattern that the two filament catalogues
are similar at high-density regions and get less similar when the density decreases.
}
\label{Fig::ex::agree_den}
\end{center}
\end{figure*}

We also study the agreement for the two catalogues at different densities.
For each point on the filaments, we assign a density value to them
by using the distance to the $1000$-th nearest 
dark matter particles (Figure~\ref{Fig::ex::agree_den}).
The $1000$ is chosen arbitrarily; we have tried other values and the result remains similar.
The inverse of this distance is an effective density
(at overdense regions, this distance is small
while at void region, this distance is large).
We then rank points on filaments according to their densities
and plot the distribution of difference under different density ranking categories.
The high-density filaments are those whose density ranking is among 
the top 10\% (red) and top 25\% (orange).
The low density filaments are those whose density ranking is in
the bottom 25\% (green) and bottom 10\% (blue).
The result is given in Figure~\ref{Fig::ex::agree_den}.
We observe
that the distribution of difference for both catalogue is small,
and the difference is even smaller for filaments
reside in high-density regions.
The reason why the difference is density-dependent is because galaxies are a biased tracer
for the matter density field so that
galaxies trace the
large-scale structure structures better at high-density regions.
Note that the result remains similar when we replace
dark matter density by galaxy density.

To conclude, Figure~\ref{Fig::ex::agree} and Figure~\ref{Fig::ex::agree_den} show that 
the filaments from galaxies and the filaments from dark matter particles
are similar and this similarity is stronger for high-density regions.
For the following analysis on the galaxy alignment,
we use the filaments from galaxies.


\subsection{Principal Axes Alignment to Filaments}


\begin{figure}
\includegraphics[width=3 in]{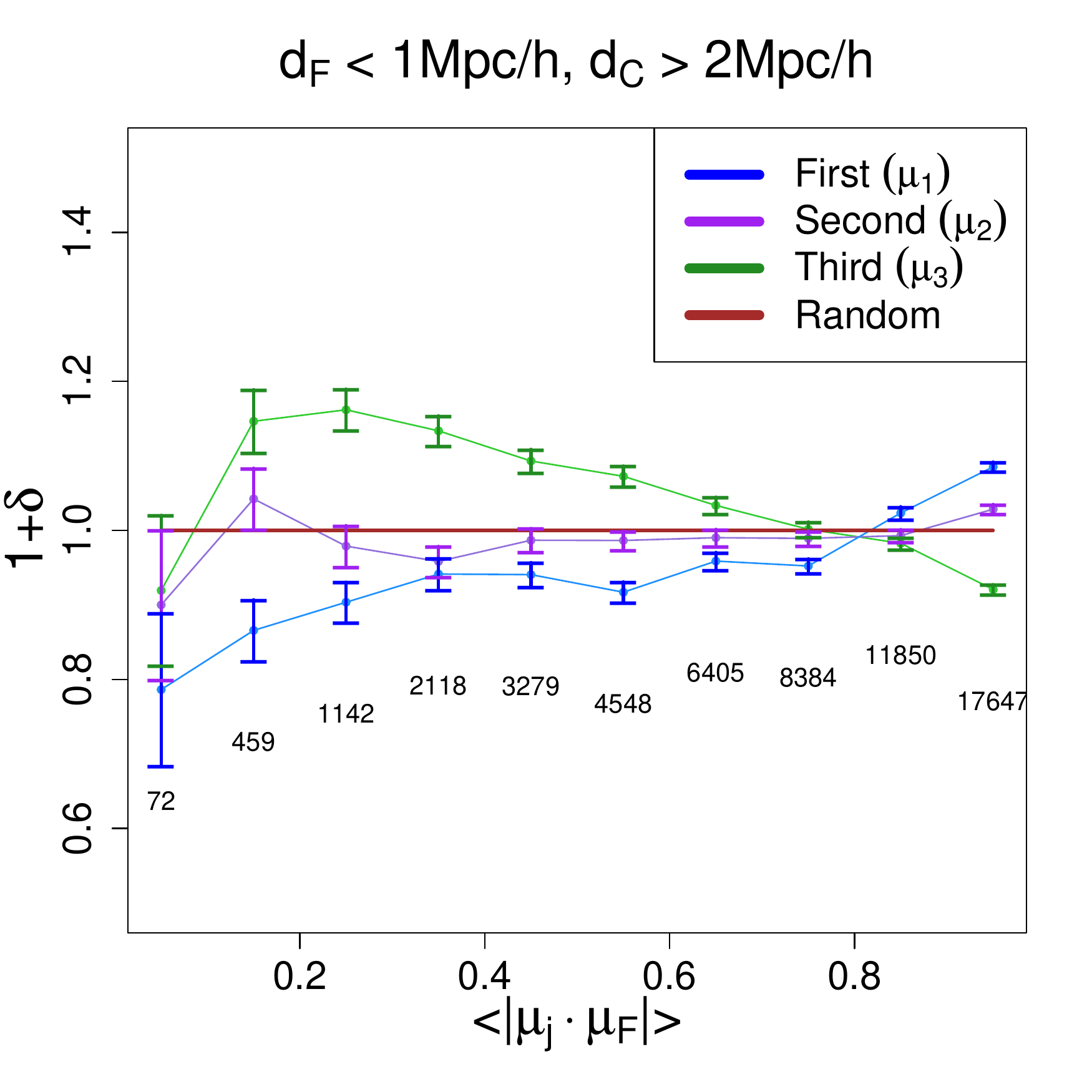}
\caption{Distribution of alignment signals $|\langle \mu_j \cdot \mu_F \rangle|$.
Note that we only consider galaxies away from groups ($d_C> 2$ Mpc$/h$) but
close to filaments ($d_F<1$ Mpc$/h$).
We observe a significant increase for $|\langle \mu_1\cdot \mu_F \rangle|\approx 1$.
This shows that the major principal axis tend to align along the orientation of
the nearby filament.
The $\delta$ in the Y-axis is the excess probability compared to the distribution of random angle.
The error bars are calculated from the bootstrap.
The number below shows the sample size
for each bin.}
\label{Fig::ang::F0}
\end{figure}

We first examine the correlation between principal axes
and $\mu_F$, the orientation of the nearest filament of a galaxy.
Note that since both $\mu_j$ and $-\mu_j$ are describing the same
axis, we take the absolute value of inner product between $\mu_j$ and $\mu_F$.

We compute the alignment signal $|\mu_j\cdot \mu_F|$
for galaxies close to filaments ($d_F<1$ Mpc$/h$) but not close to
groups ($d_C>2$ Mpc$/h$).
Note that $|\mu_j\cdot \mu_F|\approx 1$ means that the
$j$-th principal axis align along $\mu_F$ while 
$|\mu_j\cdot \mu_F|\approx 0$ indicates that they are misaligned.
In Figure~\ref{Fig::ang::F0}, 
we show the excess probability compared to the
distribution of inner product between two random unit axes.
We find that the alignment signal for first principal axis (blue curve)
is very significant (with KS-test p-value being $4\times 10^{-28}$).
By comparing the blue curve in both the left and right side of Figure~\ref{Fig::ang::F0},
we see that the distribution of the alignment is concentrated at values $\approx 1$.
This means we have more aligned galaxies compared to
two galaxies with random orientations.
The trend is reversed for the minor principal axis (with KS-test p-value $1\times 10^{-17}$).
Thus, we conclude that 
the major (first) principal axis for a galaxy tend to align along filaments
while the minor axis tends to misalign to the filaments.

\begin{figure*}
\begin{center}
\includegraphics{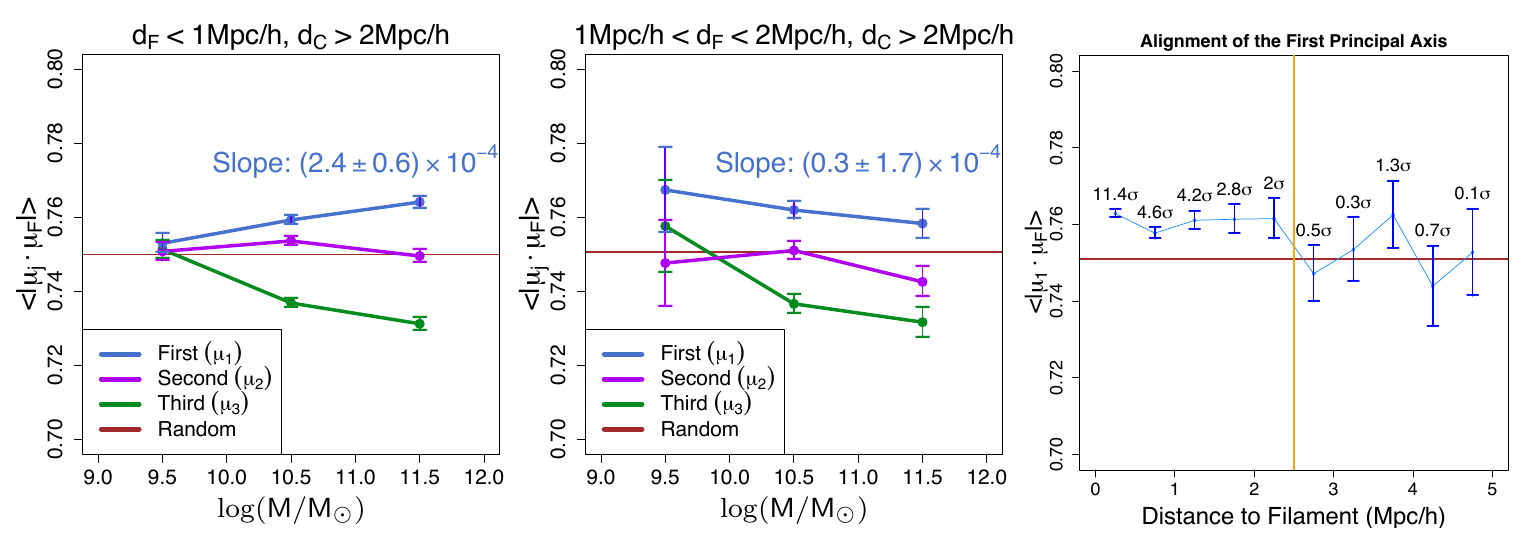}
\caption{Alignment signals $|\langle \mu_j\cdot\mu_F\rangle|$ 
versus subhalo mass and $d_F$, the distance to filaments.
Note that we remove galaxies that are
close to groups (i.e. all galaxies satisfy $d_C>2$ Mpc$/h$).
{\bf Left:}
We show how alignment signals change as a function of
subhalo mass at $d_F<$ 1 Mpc$/h$.
The slope of $|\langle \mu_1\cdot\mu_F\rangle|$ versus $d_F$ is indicated in the panel.
We observe a significant ($4\sigma$) linear effect from subhalo mass on 
the alignment of first principal axis along the nearby filament orientation.
{\bf Middle:} 
The same analysis as left panel but with different range for distance
to filaments ($1$ Mpc$/h$ $<d_F<2$ Mpc$/h$).
Compared with left panel,
we still see an alignment signal but
the mass-dependency disappears.
{\bf Right:} The alignment signal for the first principal axis,
$\langle |\mu_1 \cdot \mu_F| \rangle$, as a function of $d_F$.
Note that we display the significance on top of the alignment value for each mass bin.
The alignment signal is significant when $d_F<2.5$ Mpc$/h$ (left side of the orange line).
}
\label{Fig::ang::F}
\end{center}
\end{figure*}


To study how the subhalo mass of a galaxy affects the alignment signal,
we further partition galaxies by their mass into three mass-groups
\begin{equation}
\begin{aligned}
10^{9} M_\odot/h&<M<10^{10}M_\odot/h, \\
10^{10} M_\odot/h&<M<10^{11}M_\odot/h, \\
10^{11} M_\odot/h&<M<10^{12}M_\odot/h
\end{aligned}
\label{eq::p1}
\end{equation}
and plot the average alignment signal $\langle|\mu_j\cdot \mu_F|\rangle$
within each mass bin.
Note that the above three mass-groups contain $88\%$ of all subhalos we have 
(about $11\%$ of the subhalos have subhalo mass 
$M< 10^{9}M_\odot/h$ and $0.6\%$ have subhalo mass
$M>10^{12}M_\odot/h$).
The result is given in the first two panels in Figure~\ref{Fig::ang::F}.
The left (first) panel shows a clear patten
that the average alignment signal for major axis 
increases as the subhalo mass increases (this slope has a $4\sigma$ significance)
and the effect of misalignment for minor axis is also
augmented for massive galaxies.
\cite{2006MNRAS.370.1422A,2009ApJ...706..747Z,2013ApJ...775L..42T, 2013ApJ...779..160Z}
also report a similar mass dependency on the alignment signal.

The middle (second) panel of Figure~\ref{Fig::ang::F} shows the same analysis as
the left panel but at a different distance range to filaments.
We now focus on 
galaxies
whose distance to filaments is within $1$ Mpc$/h$ $< d_F<2$ Mpc$/h$.
That is, the middle panel shows the alignment
signal for galaxies that are mildly close to filaments (but not too close).
Our result shows that,
unlike galaxies being very close to filaments (left panel),
the alignment signal is not mass-dependent (only $0.2\sigma$).

On the right panel of Figure~\ref{Fig::ang::F},
we plot the average alignment signal for the first principal axis
at different distance bins to show how the distance
affects the alignment signal.
Note that here we only focus on galaxy with subhalo mass between 
$[10^{9}M_\odot/h, 10^{12}M_\odot/h]$.
As expected, when the distance to filaments is small (less than $2.5$ Mpc$/h$),
the alignment signal is significant ($> 2\sigma$).
However, when the distance is above $2.5$ Mpc$/h$,
we do not observe any significant alignment signal.
This result is consistent with the left and middle panels of Figure~\ref{Fig::ang::F}
since the two panels both satisfy
$d_F< 2.5$ Mpc$/h$.

\begin{figure*}
\begin{center}
\includegraphics{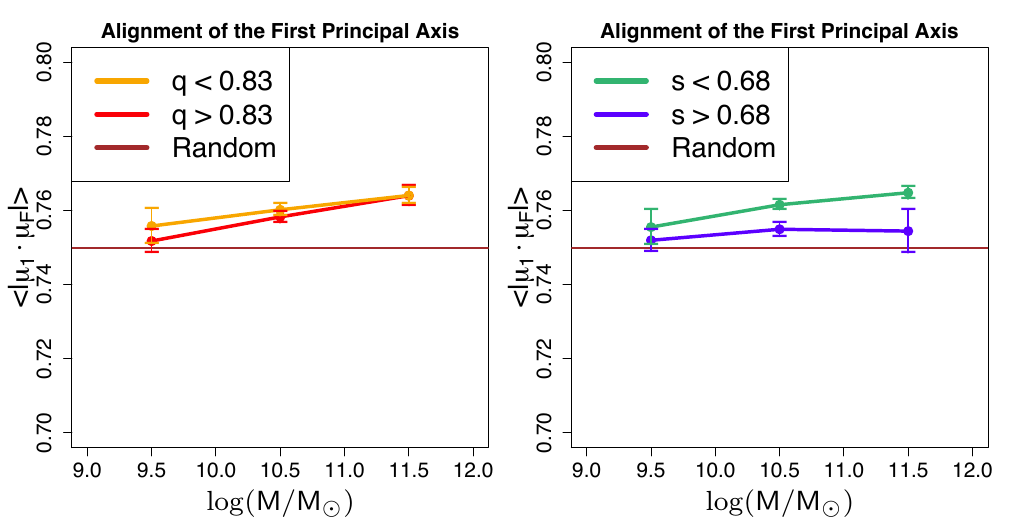}
\caption{The correlation between shapes and the alignment signal.
Note that we focus on the alignment signal of first principal axis.
We study how a galaxy's shape is correlated with
$|\langle \mu_1\cdot \mu_F \rangle|$.
We split the galaxies into groups according to their shapes, measured
by the $q$ and $s$.
The separation cuts on $q$ and $s$ are chosen by the median of $q$ and $s$ 
for galaxies with $d_F<1$ Mpc$/h$ and $d_C>2$ Mpc$/h$ .
{\bf Left:}
We separate galaxies into two groups by $q$:
more elliptical galaxies ($q<0.83$, orange) and
more spherical galaxies ($q>0.83$, red).
There is no significant difference between these two groups ($0.42\sigma$).
{\bf Right:}
We separate galaxies into two groups by $s$:
more elliptical galaxies ($s<0.68$, green) and
more spherical galaxies ($s>0.68$, blue).
In this case, we observe a significant difference between the two groups ($3.36\sigma$).
An elliptical galaxy seems to align more along the filaments compared with
a spherical one.
}
\label{Fig::ang::QS}
\end{center}
\end{figure*}

Finally, we focus on the alignment signal for first principal axis
and study how the shape ($q$ and $s$) influences the signal.
We partition galaxies according to their $q$ and $s$.
The cut on $q$ is 0.83 and the cut on $s$ is 0.68; 
both cuts are chosen by the median value for all galaxies.
The result is given in Figure~\ref{Fig::ang::QS}.
The left panel shows the alignment signal for the major axis
under the cut on $q$.
We find that the two groups do not have significant difference 
(significance is only $0.42 \sigma$).
The right panel shows the same analysis under the cut on $s$.
We see a significant difference ($3.36\sigma$) for the two groups of galaxies.
Namely, an elongated galaxy (with smaller $s$) tends to align more along nearby filaments
compared to a spherical one.
This is not an effect due to measurement error since in the simulations,
all the galaxies have $>500$ particles and the shapes and orientations are 
well-measured even in the sample that are closer to round.

\subsection{Principal Axes Alignment to Groups}	\label{sec::PAAG}

\begin{figure*}
\begin{center}
\includegraphics{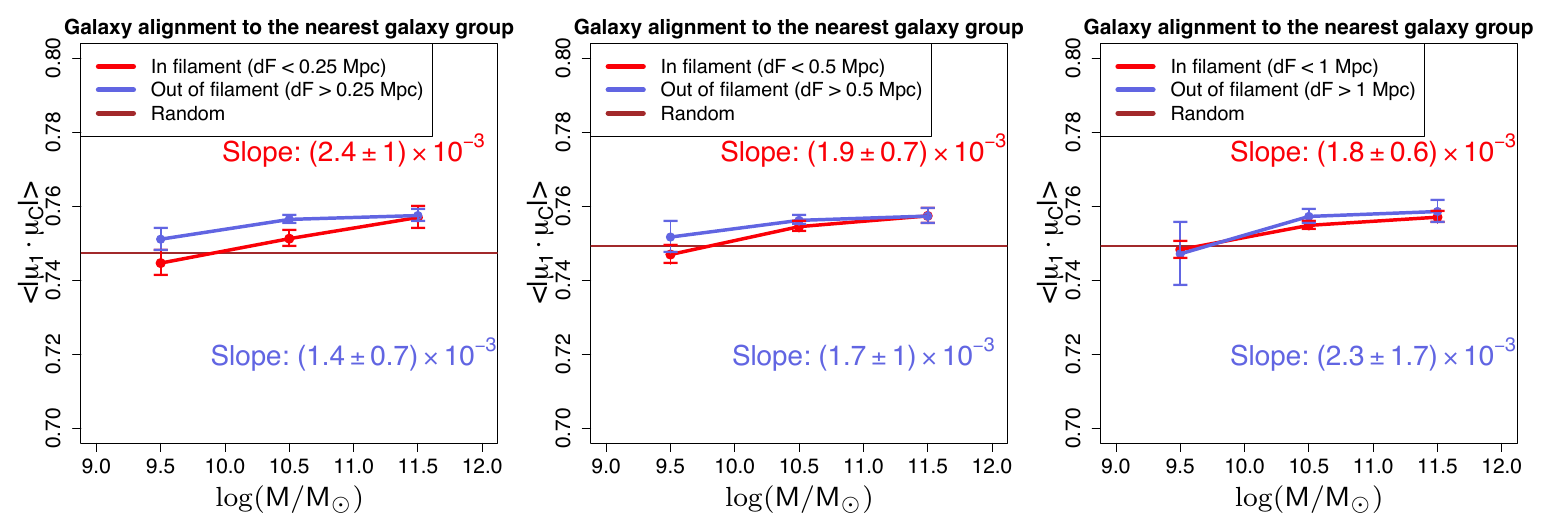}
\caption{
The effect for filaments on $|\langle \mu_1 \cdot \mu_C \rangle|$, the galaxy alignment
along `the direction toward the nearest group'.
We focus on the galaxies with $d_C>1$ Mpc$/h$
to make sure they are not inside a group.
We then partition galaxies into two types: in filaments and out filaments
by imposing a threshold $\tau$ on 
the distance to filaments (see {\S} \ref{sec::PAAG} for more details).
{\bf Left:} $\tau = 0.25$ Mpc$/h$.
Namely, a galaxy whose distance to filaments is less than $0.25$ Mpc$/h$
will be considered \emph{within filaments}.
Otherwise this galaxy is classified as \emph{out of filaments}.
{\bf Middle:} $\tau =0.50$ Mpc$/h$.
{\bf Right:}  $\tau =1.00$ Mpc$/h$.
The red color denotes $|\langle \mu_1 \cdot \mu_C \rangle|$
for galaxies that are inside a filament and the blue color
is the same but for galaxies out of filaments.
Note that we also attach the significance for the slope using
the linear regression.
}
\label{Fig::ang::GbyF}
\end{center}
\end{figure*}

We also study how filaments influence the galaxy alignment toward the nearest galaxy groups.
We focus on the inner product between the major (first) principal axis
and the direction to the nearest galaxy groups.
Namely, we test how the alignment signal $|\mu_1\cdot\mu_C|$
is different for galaxies within a filament versus outside filaments.
To classify a galaxy as being inside a filament or out of filaments,
we call a galaxy \emph{in filaments} if their distance to filament $d_F$
is less than a distance threshold $\tau$.
Otherwise, we say this galaxy is \emph{out of filaments}.
Note that we only consider galaxies that are 
at least $1.00$ Mpc$/h$ away from the nearest group
to make sure this galaxy is not in this group.

The result is given in Figure~\ref{Fig::ang::GbyF}.
We consider three different distance thresholds: $0.25$ Mpc$/h$ (left), $0.50$ Mpc$/h$ (middle),
and $1$ Mpc$/h$ (right).
In every panel, we observe that, 
for galaxies in filaments (red curves), the alignment signal significantly 
depends (the significance ranges from $2.4\sim3.0\sigma$) on the subhalo mass
for the galaxy. 
On the other hand, 
the alignments signal for galaxies outside filaments (blue curves) 
are significant only
when we set $\tau=0.25$ Mpc$/h$ but this dependency disappears for other cases.
Moreover, when we increase the threshold $\tau$,
the slope (the linear effect from mass on galaxy alignments toward groups)
for galaxies within filaments
decreases but the slope for galaxies out of filaments increases. 
This suggests that the majority of the
mass-dependency is from galaxies with $d_F< 0.25$ Mpc$/h$.

\subsection{Shape Analysis}
Now we study the correlation between 
the shape of galaxies 
and their proximity to galaxy groups and filaments.
In particular, we focus on 
$q$ for each galaxy since it measures the ratio to the length of
first and second principal axis
and is a useful summary statistic for the ellipticity.

\begin{figure*}
\begin{center}
\includegraphics{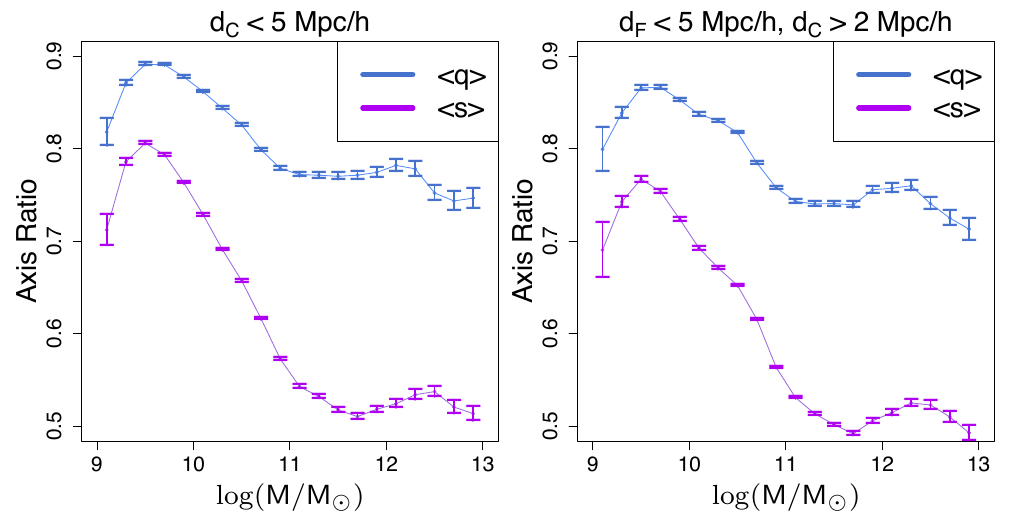}
\caption{Impact from subhalo mass on the shapes of a galaxy.
{\bf Left:} the result for galaxies close to groups ($d_C<5$ Mpc$/h$). 
{\bf Right:} the result
for galaxies close to filaments but not close to groups
($d_F<5$ Mpc$/h$ and $d_C>2$ Mpc$/h$).
The two distance ranges will be used in Figure~\ref{Fig::QS::FC}.
In both cases, we observe a strong impact on both $q,s$ from subhalo mass.
}
\label{Fig::QS::M}
\end{center}
\end{figure*}

\begin{figure*}
\begin{center}
\includegraphics{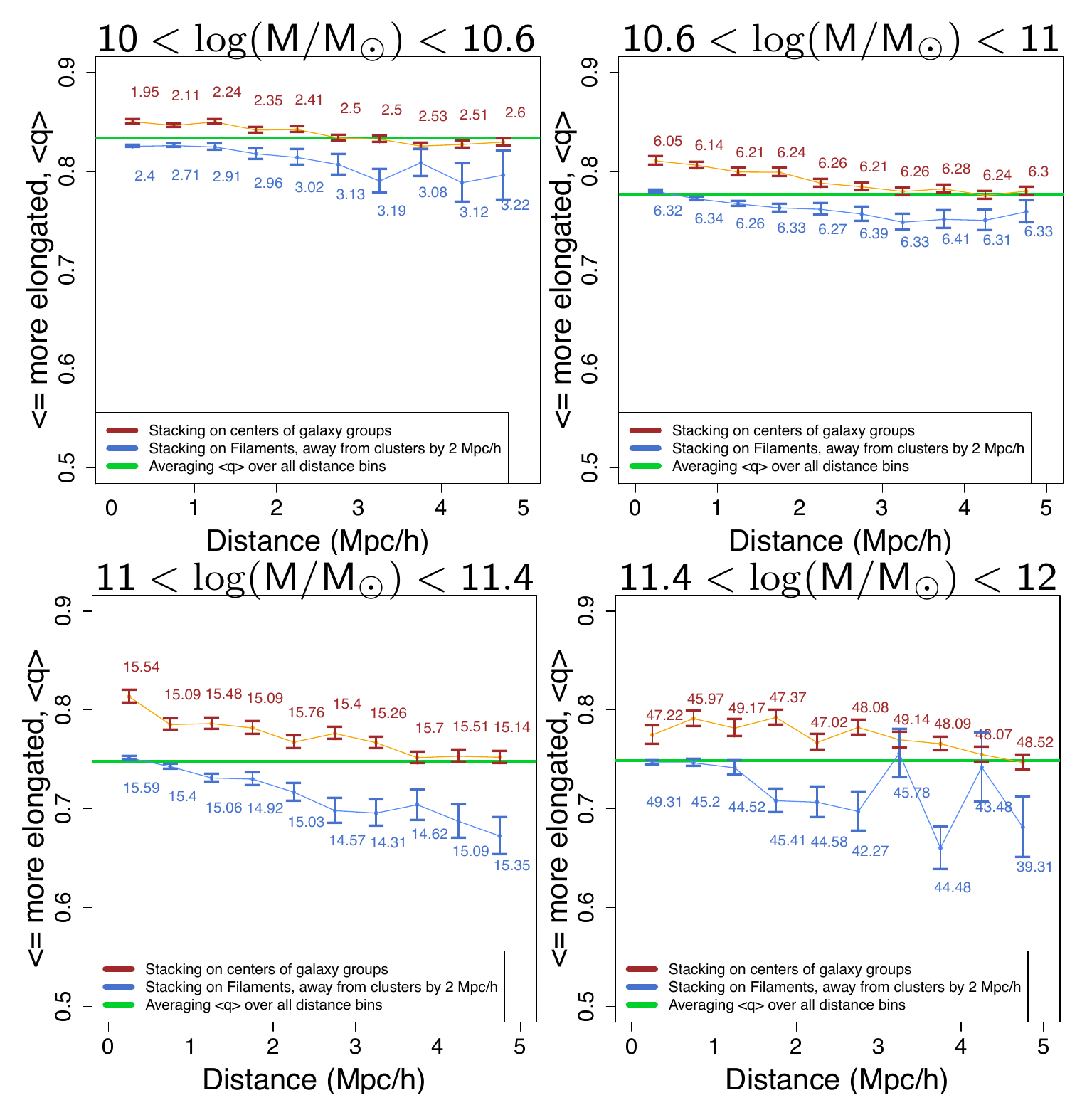}
\caption{Influence from groups and filaments on the shape of galaxies.
We display how $\langle q\rangle$ changes as the distance to filaments (blue) and groups (brown)
changes.
Note that we partition galaxies according to their subhalo mass 
to adjust the mass-dependency on $\langle q\rangle$.
The number near $\langle q\rangle$ at each bin
is the average subhalo mass within that bin.
}
\label{Fig::QS::FC}
\end{center}
\end{figure*}

We partition galaxies according to their subhalo mass into four groups:
\begin{equation}
\begin{aligned}
10^{10.0} M_\odot/h&<M<10^{10.6}M_\odot/h, \\
10^{10.6} M_\odot/h&<M<10^{11.0}M_\odot/h,\\
10^{11.0} M_\odot/h&<M<10^{11.4}M_\odot/h, \\
10^{11.4} M_\odot/h&<M<10^{12.0}M_\odot/h.
\end{aligned}
\end{equation}
Note that this partitioning is different from equation \eqref{eq::p1}
because
here we separate galaxies by subhalo mass to control the effect
from subhalo mass on $q$.
Therefore, we need a fine mass bin to reduce the mass-dependency
on $q$.
Figure~\ref{Fig::QS::M} shows how $\langle q\rangle$ and $\langle s\rangle$
change as the subhalo mass changes.
This pattern corroborates what is observed in 
\cite{2014MNRAS.441..470T}.
Note that \cite{2012JCAP...05..030S} observed a similar result
using dark matter halos and dark matter halo shapes.

Figure~\ref{Fig::QS::FC} displays
how $\langle q\rangle$ changes as $d_F$ or $d_C$ changes.
We stack galaxies around filaments (and groups)
and then partition galaxies according to their distance to filaments (and groups)
into 10 bins with a bin width of $0.5$ Mpc$/h$ 
(so we only consider distance to filaments/groups less than $5$ Mpc$/h$).
For galaxies stacking around filaments,
we have an additional constraint $d_C > 2$ Mpc$/h$
to eliminate the effect from galaxy groups.
For each bin, we also show the average subhalo mass.

Overall, we find two cases that lead to a change in the $\langle q \rangle$
after adjusting the effect from subhalo mass.
The first case can be found at the top right and bottom left panel of Figure~\ref{Fig::QS::FC}. 
We observe a decreasing pattern
as $d_F$ or $d_C$ increases.
In other words, for galaxies close to filaments or groups,
we expect their $q$ to be higher (more rounded galaxies).
In both panels, the subhalo mass within each bin does not change too much
so that this effect cannot be explained by the mass difference.
The second case is on the difference between groups
and filaments.
In every panel of Figure~\ref{Fig::QS::FC},
we find that, for galaxies at the same 
distance to either filaments or groups,
$\langle q\rangle$ is always higher for galaxies close to groups.

An explanation to the above cases is due to the difference 
in the ratio of elliptical galaxies.
If elliptical galaxies are more likely to be in/near groups and filaments
and if they are usually rounder than disk galaxies,
then we would expect to see that galaxies close to groups and filaments
are on average rounder than those that are far away.
And if this effect is stronger for groups (perhaps due to a stronger tidal field),
then we should see a higher $q$ for galaxies that are closer to groups.


Note that in the upper left panel ($10^{10.0} M_\odot/h<M<10^{10.6} M_\odot/h$), 
we find a decreasing trend
for the average subhalo mass when the distance to large-scale
structure is increasing.
Moreover, at the same distance,
galaxies close to groups tend to be lower mass
compared with those close to filaments.
Namely, one will find more low-mass
galaxies (with subhalo mass less than $10^{10.6} M_\odot/h$)
at strong tidal force regions.
We do not find a similar patten in other mass range, so this 
phenomena occurs only for light galaxies.
A possible reason is that,
in a region with strong tidal effect,
matter concentrates around the large-scale structures
so that there will be many newly formed galaxies.
These young galaxies are generally light, which decreases
the average mass.

\section{Conclusion}	\label{sec::conc}
In this paper, we study the correlation between galaxy orientation and filaments
by applying the density ridge model to the smoothed particle hydrodynamic simulation.
A short summary for what we have found is as follows:
\begin{enumerate}
\item We demonstrate that filaments constructed using density ridges 
from galaxy density field
are similar to filaments constructed from dark matter density field.
This analysis,
along with the fact that density ridges are statistically consistent
in both theory
\citep{2014arXiv1406.5663C} and simulation \citep{2015arXiv150105303C},
suggests that we might be able to reconstruct filaments 
of dark matter field by using only galaxies.
However, this simulation goes down to quite low subhalo mass compared to most observations.
More analysis needs to be done in future for real observations.

\item We find that galaxies align along with their nearby filaments
and this alignment is stronger for massive galaxies when the galaxy
is close to filaments ($d_F<1$ Mpc$/h$).
This corroborates the past literature
\citep{2006MNRAS.370.1422A,2009ApJ...706..747Z,2013ApJ...775L..42T, 2013ApJ...779..160Z}.
Moreover, we discover that the alignment signal 
has a sphere of influence of $ 2.5$ Mpc$/h$ in radius.
Namely, the principal axes of a galaxy whose distance to
the nearest filaments is less than $2.5$ Mpc$/h$
would be influenced by the nearby filaments. 
This critical radius has not been previously reported in the literature.

\item We also observe an interesting pattern: 
the galaxy alignment toward the nearest 
galaxy group is
mass-dependent when 
this galaxy is near a filament within $0.25$ Mpc$/h$.
For galaxies close to a filament, massive galaxies tend to align along 
the direction toward the nearest group.
For galaxies away from filaments, subhalo mass has little impact on the alignment 
toward a group.

\item Finally, we find that the ellipticity of a galaxy is correlated with
its distance to groups and filaments when the subhalo mass 
is between $10^{10.6}M_\odot$ and $10^{11.4} M_\odot$.
A galaxy that is close to either a group or a filament tends to be 
less elliptical (more rounded).
\end{enumerate}

The galaxy-filament alignment observed in this paper can be explained
by the intrinsic alignment model and the relationship between tidal field and filaments.
The intrinsic alignment model links the galaxy principal axes
to the tidal field
\citep{2001MNRAS.320L...7C, 2004PhRvD..70f3526H, 2010PhRvD..82d9901H}
and the tidal field is often explicitly or implicitly used 
to define a filament 
\citep{2007MNRAS.375..489H,2007MNRAS.381...41H,2009MNRAS.396.1815F, Cautun2012}.
Therefore, the principal axes and the shape of a galaxy are expected to
be correlated with its nearby filament.
Based on the galaxy-filament interaction we have found, 
we expect the future survey on intrinsic alignment and 
weak lensing to be improved if one incorporates the effect from filaments.
This might be a way to extend the current linear or non-linear alignment model 
\citep{2007NJPh....9..444B, 2011JCAP...05..010B, 2015arXiv150402510B}
since the alignment along filaments can be viewed as a new type of non-linear effect.

\section*{Acknowledgments}
This work is supported in part by the Department of Energy under grant DESC0011114; 
YC is supported by DOE;
SH is supported in part by DOE-ASC, NASA and NSF; 
AT was supported for the duration of this work by NASA ROSES 12-EUCLID12-0004.  
RM was supported in part by NSF under
Grant. No. AST-1313169 and NASA ROSES 12-EUCLID12-0004.
CG is supported in part by DOE and NSF;
LW is supported by NSF.


\bibliographystyle{mnras}
\bibliography{SuRF_IA.bib}
\bsp

\label{lastpage}

\end{document}